\documentclass[usenatbib]{mnras}
\usepackage[utf8]{inputenc}
\usepackage{amsmath}
\usepackage{graphicx}
\usepackage[T1]{fontenc}
\usepackage{ae,aecompl}
\usepackage{subfig}

\newcommand{\cm}{{\mathrm{cm}}}
\newcommand{\km}{{\mathrm{km}}}
\newcommand{\s}{{\mathrm{s}}}
\newcommand{\cmsqps}{\cm^2\s^{-1}}
\newcommand{\pc}{{\mathrm{pc}}}
\newcommand{\Myr}{{\mathrm{Myr}}}
\newcommand{\K}{{\mathrm{K}}}
\newcommand{\Msun}{\mathrm{M}_{\odot}}

\title[The angular momentum structure of cosmic ray driven galactic outflows triggered by stream accretion]{The angular momentum structure of cosmic ray driven galactic outflows triggered by stream accretion}

\author[N. Peschken, M. Hanasz, T. Naab, D. W\'{o}lta\'{n}ski, A. Gawryszczak]{N. Peschken$^{1}$\thanks{Contact e-mail:
    \href{mailto:npeschken@umk.pl}{npeschken@umk.pl}}, M. Hanasz$^{1}$, T. Naab$^{2}$, D. W\'{o}lta\'{n}ski$^{1}$, A. Gawryszczak$^{3}$\\
$^{1}$ Institute of Astronomy, Faculty of Physics, Astronomy and Informatics, Nicolaus Copernicus University,\\
ul. Grudziadzka 5/7, PL-87-100 Toruń, Poland\\
$^{2}$ Max Planck Institute for Astrophysics, Karl-Schwarzschild-Str. 1, 85748 Garching, Germany\\
$^{3}$Nicolaus Copernicus Astronomical Center, Bartycka 18, 00-716 Warsaw, Poland
}

\pubyear{2020}

\begin{document}
\label{firstpage}
\pagerange{\pageref{firstpage}--\pageref{lastpage}}
\maketitle

\begin{abstract}
  We investigate the impact of gas accretion in streams on the evolution of disc galaxies, using magneto-hydrodynamic simulations including advection and anisotropic diffusion of cosmic rays generated by supernovae as the only source of feedback. Stream accretion has been suggested as an important galaxy growth mechanism in cosmological simulations and we vary their orientation and angular momentum in idealised setups. We find that accretion streams trigger the formation of galactic rings and enhanced star formation. The star formation rates and consequently the cosmic ray driven outflow rates are higher for low angular momentum accretion streams, which also result in more compact, lower angular momentum discs. The cosmic ray generated outflows show a characteristic structure. At low outflow velocities (< 50 km/s) the angular momentum distribution is similar to the disk and the gas is in a fountain flow. Gas at high outflow velocities (> 200 km/s), penetrating deep into the halo, has close to zero angular momentum, and originates from the centre of the galaxies. As the mass loading factors of the cosmic ray driven outflows are of order unity and higher, we conclude that this process is important for the removal of low angular momentum gas from evolving disk galaxies and the transport of, potentially metal enriched, material from galactic centres far into the galactic haloes.
\end{abstract}
 
\begin{keywords}
galaxies: evolution -- galaxies: kinematics and dynamics -- galaxies: spiral -- galaxies: structure
\end{keywords}

\section{Introduction}

Cosmic rays (CRs) are high energy particles, mostly protons and electrons travelling at relativistic speeds, which are thought to be produced in supernova remnants by diffusive shock acceleration (\citealt{1977DoSSR.234.1306K, 1978ApJ...221L..29B, 2004MNRAS.353..550B}). Their presence in galaxies (with an energy density approximately in equipartition to the thermal gas and magnetic energy densities, see e.g. \citealt{1990ApJ...365..544B}) can have a significant impact on the ISM, in particular through the dynamical coupling between gas and CRs, which allows CR energy and momentum to be transferred to the gas. Being a non thermal component (\citealt{2013PhPl...20e5501Z,2015ARA&A..53..199G}), CRs do not cool as quickly as thermal gas, and therefore constitute a long term energy reservoir while travelling large distance across the galaxy via advection and diffusion mechanisms (\citealt{2007ARNPS..57..285S,2013PhPl...20e5501Z,2015ARA&A..53..199G}), preferably along the magnetic field lines (\citealt{1999ApJ...520..204G}). This allows cosmic rays to escape the galactic disc in spiral galaxies, creating a steady gradient of cosmic ray pressure. The presence of this gradient supports gas outflows from the disc (\citealt{1969ApJ...156..445K,1993A&A...269...54B, 2008ApJ...674..258E,2012A&A...540A..77D}), and launches powerful galactic-scale winds (\citealt{1991A&A...245...79B, 2012MNRAS.423.2374U, 2013ApJ...777L..38H, 2013ApJ...777L..16B, 2014ApJ...797L..18S, 2019A&A...630A.107D}). These cosmic-ray driven outflows have been studied in some recent simulations (e.g. \citealt{2014MNRAS.437.3312S, 2017ApJ...834..208R,2017MNRAS.467..906W, 2019MNRAS.488.3716C, 2020MNRAS.497.1712B, 2020MNRAS.492.3465H}), but the inclusion of CRs together with magnetic field in galaxy simulations is still not very common in galaxy simulations and requires more investigation. From recent work, it seems that the presence of cosmic rays, by driving powerful galactic outflows, thickens galactic discs (\citealt{2014MNRAS.437.3312S, 2016ApJ...816L..19G}), suppresses star formation with the removal of gas from star forming areas (\citealt{2017ApJ...834..208R, 2020MNRAS.492.3465H}), and triggers galactic fountain mechanisms (e.g. \citealt{2018ApJ...856..112F}). Cosmic rays coupled with magnetic field therefore appear to have a strong influence on galaxy evolution, and are critical to understand gas dynamics in gas-rich galaxies (see \citealt{2017ARA&A..55...59N} for a review). \\
The angular momentum structure of outflows is of particular importance as outflows regulate the angular momentum of disc galaxies, re-distribute their gas, and seem to feed predominantly from low angular momentum gas (\citealt{2010AdSpR..46..485D,2011MNRAS.415.1051B,2014MNRAS.443.2092U,2015ApJ...804L..40G, 2017ApJ...841...16D}). Most previous simulations used to study angular momentum in outflows did not investigate outflows generated by cosmic rays alone to clearly isolate this driving mechanism from energy or momentum driven outflows. In this paper, we study the link between gas angular momentum and cosmic ray driven outflows, using the MHD code \textsc{piernik}. Cosmological simulations of galaxies and structure formation predict that gas can be accreted from the cosmic web in the form of streams (e.g. \citealt{2009Natur.457..451D}). It has been suggested that there is a connection between the theoretically predicted stream accretion and the original observation (\citealt{2008ApJ...687...59G}) that many high redshift galaxies show star forming rings (\citealt{2012ApJ...745...11G,2015MNRAS.449.2087D,2020MNRAS.496.5372D}). We mimic this accretion process by setting up a star forming disc galaxy accreting gas in streams in the disc plane and off the disc plane with varying angular momentum, and observe the resulting generated outflows. We include only cosmic ray feedback and no thermal feedback to isolate effect of cosmic rays on outflows, as was done in \cite{2013ApJ...777L..38H}. We will first introduce our simulations in section \ref{presentation}, describing the \textsc{piernik} code and the setup for our simulations, and will then present our results in section \ref{results}, where we describe how outflows are generated and investigate the impact of angular momentum on them, as well as their AM properties and influence on the disc. We discuss our results in section \ref{discussion}, and conclude in section \ref{ccl}.

\begin{figure}
  \includegraphics[scale=0.5]{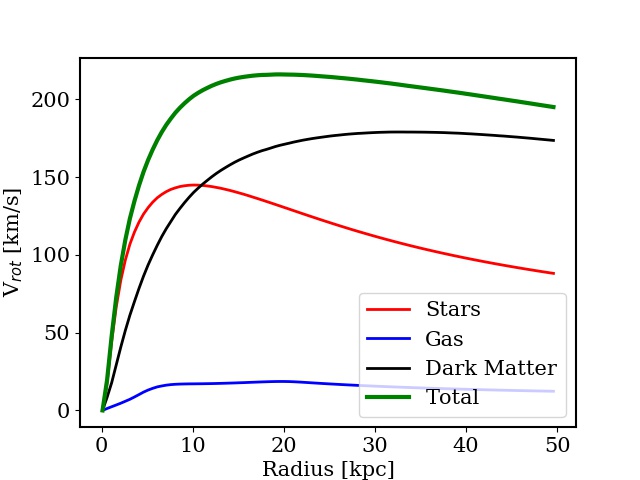}
\caption{
  Circular velocity curve (green) of the initial disc galaxy model separated into the contribution from the disc stars (red), gas (blue), and dark matter (black).}
  \label{rotcurv}
\end{figure}

\begin{figure*}
  \includegraphics[scale=0.52, trim=80 0 0 0, clip ]{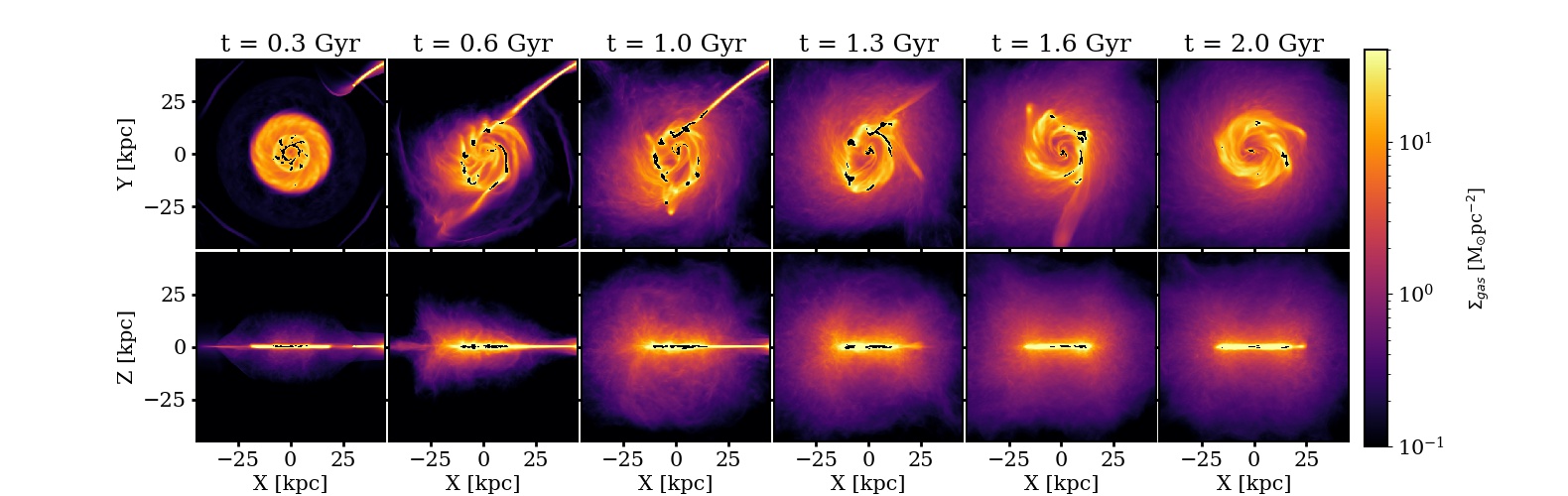}
\caption{Time evolution of the gas column density, at six different times (from left to right) for the p1 simulation: t = 300~Myr, 600~Myr, 1~Gyr, 1.3~Gyr, 1.6~Gyr, and 2~Gyr. The top panels display the face-one views, while the bottom panels show the edge-on views. The dark patches in regions of high gas density (orange/yellow) indicate areas of active star formation. We start from a stellar disc with gas, receiving an inflow of gas in the form of a co-rotating stream entering in the plane of the disc for 1~Gyr. The edge-on panels show the cosmic ray driven gas outflow from the disc to the halo.}
  \label{p1evol}
\end{figure*}

\section{Presentation of the simulations}
\label{presentation}

\subsection{The PIERNIK code}
\label{piernik}

To run our simulations, we use the \textsc{piernik} code (\citealt{2010EAS....42..275H, 2010EAS....42..281H, 2012EAS....56..363H, 2012EAS....56..367H}, a grid-MHD code based on the Relaxing Total Variation Diminishing (RTVD) scheme  \citep{jin-xin-95,2003PASP..115..303T,2003ApJS..149..447P}.
The functionality of \textsc{piernik} includes the modeling of multiple fluids: gas, dust, cosmic rays, and their gravitational and electromagnetic interactions. \textsc{piernik} is parallelized on the base of MPI library, and its data I/O communication utilizes parallel HDF5 output. \textsc{piernik} is available from the public git repository linked to the code webpage: \href{http://piernik.umk.pl}{piernik.umk.pl}.

Thermal gas and CRs are simulated as two separate fluids interacting within the fluid approximation (\citealt{2003A&A...404..389H}). We give here a brief summary of the code, and refer the reader to the four papers mentioned above, as well as to \cite{2013ApJ...777L..38H}, for a more detailed description. \\
The system of resistive MHD equations is assumed with an isothermal equation of state for the thermal gas component at $T = 10000$~$\K$ and a uniform magnetic diffusivity $\eta = 3 \times 10^{25}\cmsqps$.
The induction equation is solved with a constraint transport algorithm (\citealt{1988ApJ...332..659E}). Gravity is handled with a multigrid solver \citep{doi:10.1137/S1064827598346235} for the Poisson equation, combined with a multipole solver the potential at the domain boundaries, following \citet{1977JCoPh..25...71J} for our boundary cells implementation.\\
CRs propagate anisotropically with the diffusion-advection transport equation, moving preferentially along magnetic field lines:

\begin{equation}
\frac{\partial e_{cr}}{\partial t} + \pmb{\nabla}(e_{cr}\pmb{v}) = \pmb{\nabla}(\hat{K}\pmb{\nabla}e_{cr}) - p_{cr}(\pmb{\nabla}\cdot\pmb{v}) + Q
\end{equation}
with $e_{cr}$ being the cosmic ray energy density, $\hat{K}$ the diffusivity coefficient, $p_{cr}=(\gamma_{cr}-1)e_{cr}$ the cosmic ray pressure (we use $\gamma_{cr}=4/3$), $\pmb{v}$ the gas velocity, and $Q$ the cosmic ray source term from supernova remnants.
We take diffusion coefficients of $K_\parallel=9 \times 10^{28}\cmsqps$ in the direction parallel to the magnetic field, and $K_\perp = 9 \times 10^{26} \cmsqps$ in the perpendicular direction. These coefficients are in agreement with recent work, investigating different values for the different coefficients and their impact on galaxy evolution and outflows (\citealt{2016ApJ...816L..19G, 2016MNRAS.456..582S, 2016ApJ...827L..29S, 2018MNRAS.479.3042G,2019MNRAS.488.3716C, 2019A&A...632A..12S, 2020MNRAS.492.3465H}), finding values of the order of $10^{29} \cmsqps$ to be both realistic and generating galactic scale outflows.  \\

To model stars and dark matter, we use Nbody particles, a new addition to \textsc{piernik}. The integration is done with the leapfrog scheme together with particle mesh, projecting particle density to the grid using the Triangular Shaped Cloud (TSC) scheme. The resulting gravitational potential is then computed on the grid using the multigrid Poisson solver, with use of the multipole solver to handle particles outside the domain. The TSC algorithm is used again to derive the particles accelerations.The simulations presented in this paper also constitute a good test for the new particle solver, which proves successful as we find a realistic evolution of the stellar disc. As shown later in Fig.~\ref{evolpart}, the disc stays thin over time and develops a spiral structure.

   \subsection{Initial conditions for a disc galaxy}
   \label{disc}

   We use a setup for a disc galaxy similar to \cite{2013ApJ...777L..38H}. \\
   To build our disc galaxy, we start from a stellar disc and a dark matter halo represented by Nbody particles as described in section~\ref{piernik}, and build the gaseous disc in equilibrium with those components. \\
   We use one million particles in our simulations, half for the stellar disc component, and the other half for the dark matter halo. Those components were created with the \textsc{buildgal} software (\citealt{1993ApJS...86..389H}), using the following parameters: the stellar disc is of mass 8.56 $\times$ $10^{10}$~$\Msun$ with 5.3~kpc scale-length, 1.6 kpc scale-height and 79.5~kpc cut-off radius, and the isothermal halo has a mass of 4.28 $\times$ $10^{11}$~$\Msun$, with a core radius of 5.3~kpc and a cut-off radius of 265~kpc. This middle-range halo mass is in accordance with later results suggesting that massive haloes above $\sim 10^{12}$~$\Msun$ prevent the formation of cosmic ray driven outflows (\citealt{2018MNRAS.475..570J,2018MNRAS.477..531F}). The disc is exponential in radial direction with a hyperbolic secant function to model the vertical distribution, and a Toomre parameter of 1.5, while the halo is isothermal with a core density profile. \\
   Once we have the stellar disc and dark matter halo distribution, we compute the corresponding gravitational potential in \textsc{piernik} with the Poisson solver, and use it to build the gaseous disc in hydrostatic equilibrium. Thereafter we add gas to the disc through an external stream, as will be described in section \ref{stream}. \\
   We show the initial circular velocity curve in Fig.~\ref{rotcurv} for the different components of our simulations.
   The gas disc is initialised with a position dependent magnetic field in pressure equipartition with gas with an initial toroidal strength of 9.6~$\mu$G, and assume cosmic rays are absent in the initial state, as they will be produced throughout the simulation in supernovae explosions. \\
   Each simulation is run for 2~Gyr in a (100~kpc)$^3$ box, with a resolution of 512$^3$ cells, the disc being placed at the centre of the domain, in the XY plane. \\
   Our gaseous disc includes a model for star formation, whose star formation rate is derived with: \\
\begin{equation}
\rho_{SFR} = \epsilon \sqrt{\frac{G\rho^3}{32\pi}}
\end{equation}
provided that gas density exceeds some threshold density $\rho_{\mathrm{thr}}$ = 0.035~$\Msun$~$\pc^{-3}$ (which represents 1 Hydrogen atom~cm$^{-3}$ + 1 Helium atom every 10 H atoms) which we treat as a free parameter,
with $\epsilon$ being the SFR efficiency, set here to 10\%. 
The stellar feedback in our simulations consists in cosmic ray energy injection only; we assume that a supernova occurs for every 100~$\Msun$ of star forming gas, and that 10\% of its energy accelerating cosmic rays. Note that we do not include any actual star formation (no stellar particle creation), which seems a reasonable approximation for a 2~Gyr run time as the typical star formation efficiency is of the order of a few percents.

\begin{figure*}
  \includegraphics[scale=0.7, trim=20 0 0 0]{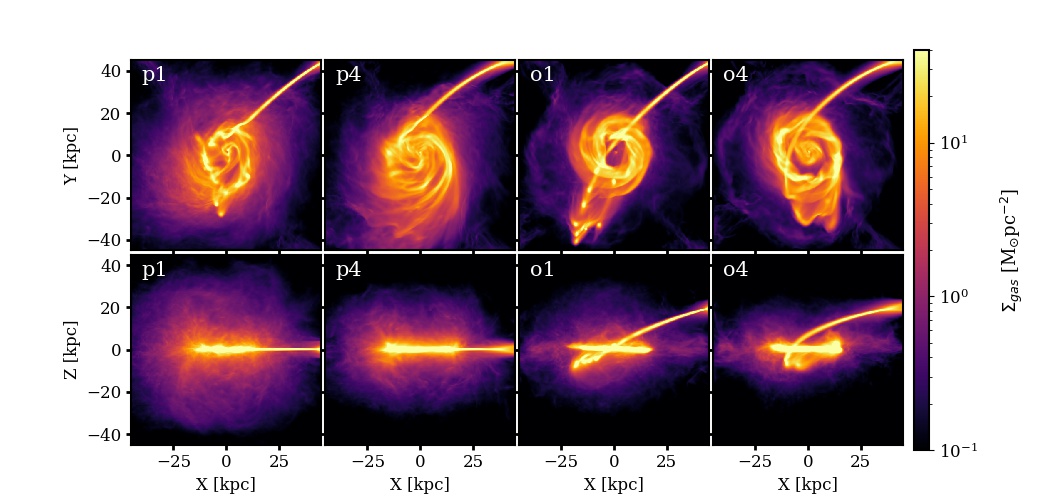}
\caption{Face-on (top panels) and edge-on (bottom panels) views of the gas column density for 4 of our simulations at t = 1~Gyr: from left to right p1 (lowest AM input, in-plane stream), p4 (highest AM input, in-plane stream), o1 (lowest AM input, out-of-plane stream), o4 (highest AM input, out-of-plane stream). The simulations p2, p3, o2 and o3 (not shown here) constitute intermediate cases of AM input with the same setup.}
  \label{setup}
\end{figure*}

      \subsection{Angular momentum input through a gas stream}
      \label{stream}

   To study the impact of inflows with varying angular momentum on cosmic ray driven outflows, we investigate what happens when a galaxy absorbs gas from the intergalactic medium, and vary the angular momentum of this accreted gas. This could represent the stream feed growth of galaxies as described in \citet{2009Natur.457..451D}, or the accretion of a gas cloud, and allows us to start with the same galaxy receiving different angular momentum inputs, to compare the resulting outflows.\\
   We model this accretion as a form of a stream starting from the edge of the domain, at the YZ side (Z being the vertical direction of the disc) at Y = 45~kpc, and having a velocity in the X direction towards the inside of the domain. We display the evolution of one simulation in Fig.~\ref{p1evol} to show our setup, with the gas stream being accreted by the disc. To have a better control over the injected angular momentum, we assume no star formation (and therefore no cosmic ray injection) in the stream until it reaches the disc. Just varying the value of the inflow velocity then allows us to vary the amount of angular momentum that will be added to the disc, as will be shown in Section \ref{comp}. To ensure that the amount of gas brought to the disc is the same in every case, we keep the flux of inflowing gas constant ($10^7$~$\Msun$~$\Myr^{-1}$) in the different simulations, by only changing the density of the stream in accordance to its velocity, while keeping its cross section constant at a value of (5~kpc)$^2$. We maintain this constant gas inflow for 1~Gyr from the start of the simulation (for a total gas mass added of 10$^{10}$~M$_{\odot}$), so that by the end of the simulation (t = 2~Gyr) the disc has absorbed all the stream and had enough time to stabilize afterwards. \\
   We run 8 simulations, each with the same initial disc, but with a different inflowing stream. Four simulations are run with the inflow of external gas arriving in the disc plane, with $Z=0$, and four with the inflow above the disc plane, with $Z=20$~kpc. In each case the four simulations correspond to a different value of the initial X-velocity of the stream (negative because the stream flows into the domain from the right side): $-15$, $-25$, $-50$ and $-75$ $\pc$~$\Myr^{-1}$ ($1$~$\pc$~$\Myr^{-1}$ = 1.023 $\km$~$\s^{-1}$). We name those 8 simulations respectively p1, p2, p3, p4 for the in-plane streams, and o1, o2, o3 and o4 for the out-of-plane ones. We refer the reader to Fig.~\ref{setup} for a visualization of 4 of those simulations and their differences in the stream input. In our setup, the higher the velocity of the stream in absolute value is, the more angular momentum it brings to the disc. We also ran an additional simulation without inflowing stream as a reference run, which we will refer to as NS (no stream).\\
   We also refer the reader to online videos of the gas evolution for all our simulations, which can be found on the \textsc{piernik} wepbage (\href{http://piernik.umk.pl/results/2021b/}{piernik.umk.pl/results/2021b}).\\

   \section{Results}
   \label{results}

   \subsection{Evolution and creation of outflows}
   \label{p1}

\begin{figure*}
  \includegraphics[scale=0.59, trim=50 0 0 0, clip]{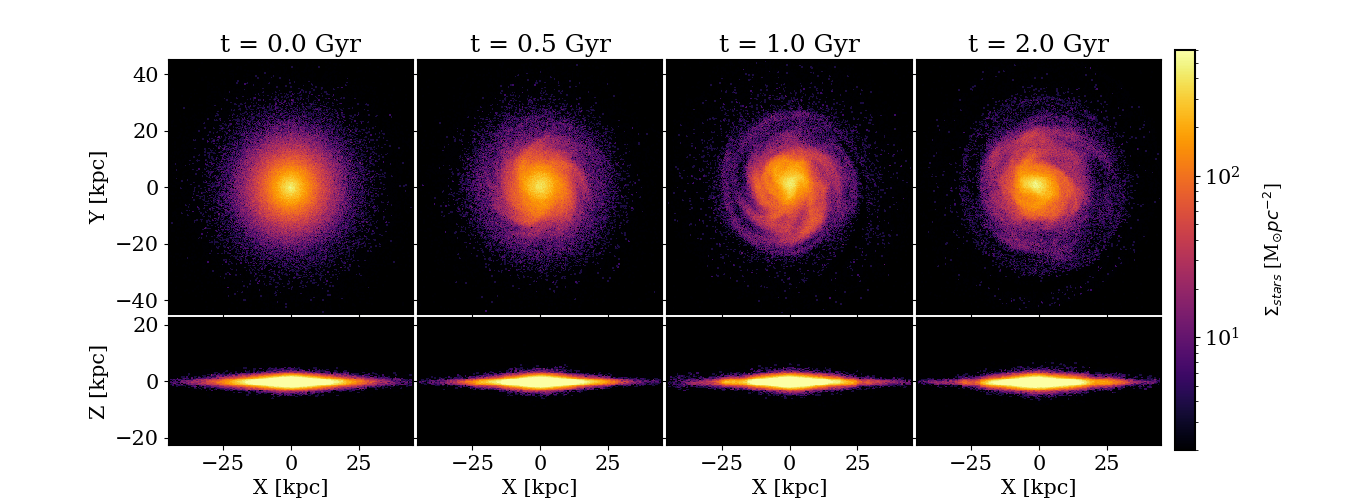}
\caption{Stellar projected density for the p1 simulation seen face-on (top row) and edge-on (bottom row) at t = 0, 0.5~Gyr, 1~Gyr and 2~Gyr.}
  \label{evolpart}
\end{figure*}

\begin{figure*}
  \includegraphics[scale=0.5, trim=80 0 0 0, clip]{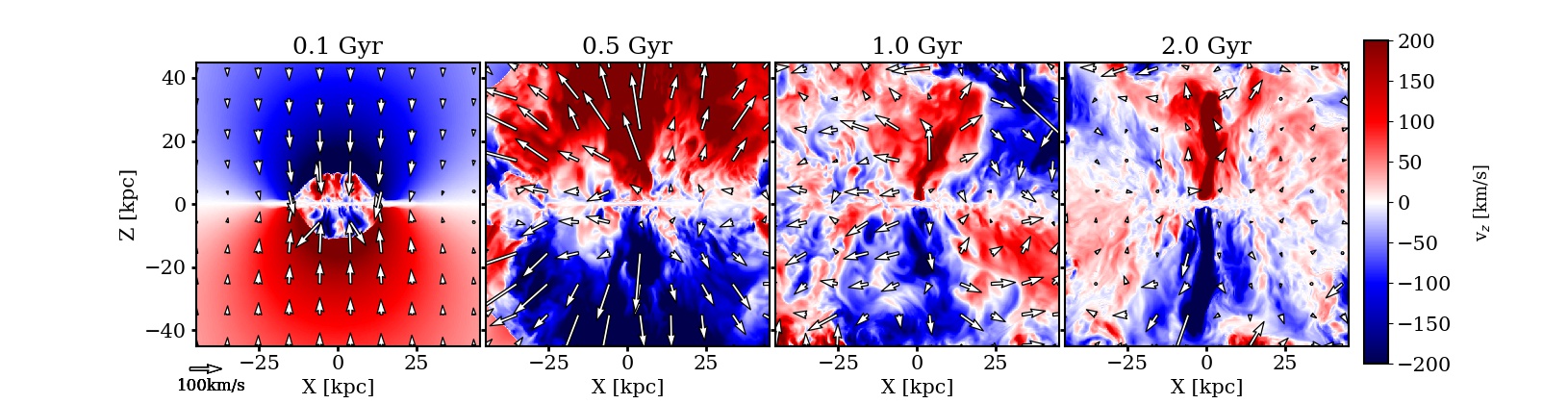}
\caption{Vertical velocity maps in the edge-on view (slice at Y = 0) for the p1 simulation, at t = 100~Myr, 500~Myr, 1~Gyr and 2~Gyr. The arrows represent the corresponding gas motion in the XZ plane. While at the start, gas vertical motions are dominated by the accretion of halo gas towards the disc, from t $\sim$ 0.5~Gyr outflows are the dominating mechanism, with strong outflows from the disc. At later times inflows grow stronger and seem to balance approximately the amount of outflows (as shown later in Fig.~\ref{inf_outf}).}
  \label{evolvz}
\end{figure*}

\begin{figure}
  \includegraphics[scale=0.55]{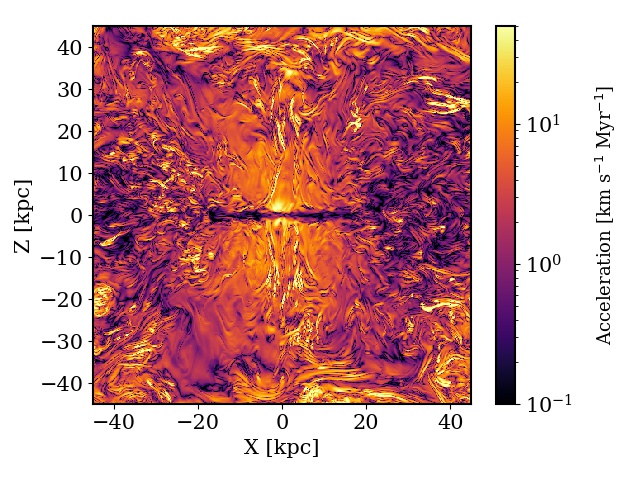}
\caption{Vertical cosmic ray gradient acceleration at 2~Gyr for the p1 simulation seen edge-on, in a slice at Y = 0. The acceleration is very strong close to the disc, especially near the centre.}
  \label{cracc}
\end{figure}

\begin{figure}
  \includegraphics[scale=0.5]{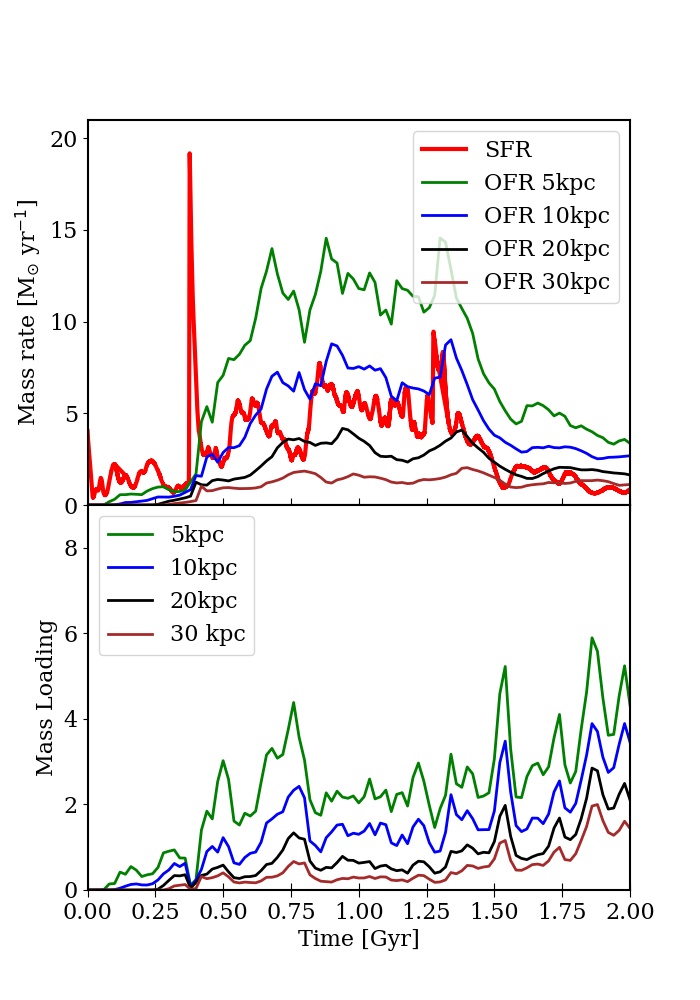}
\caption{Top panel: vertical outflow rate (OFR) for the p1 simulation as a function of time, at different altitudes above (and below) the disc, alongside the star formation rate (red) of the galaxy. Strong star formation occurs when the inflowing stream hits the disc, and remains high as long as the latter is being fed by the stream. The outflows are following the SFR closely, being directly linked to the cosmic rays injected in star forming regions. Bottom panel: corresponding mass loading as a function of time, i.e. the ratio of OFR to SFR, at different distances from the disc. The mass loading is of order of unity at 10~kpc from the disc, but increases towards the end due to the very low star formation rate, while the CR gradient of pressures stills allows for outflows.}
  \label{outfp1}
\end{figure}

\begin{figure*}
  \includegraphics[scale=0.6]{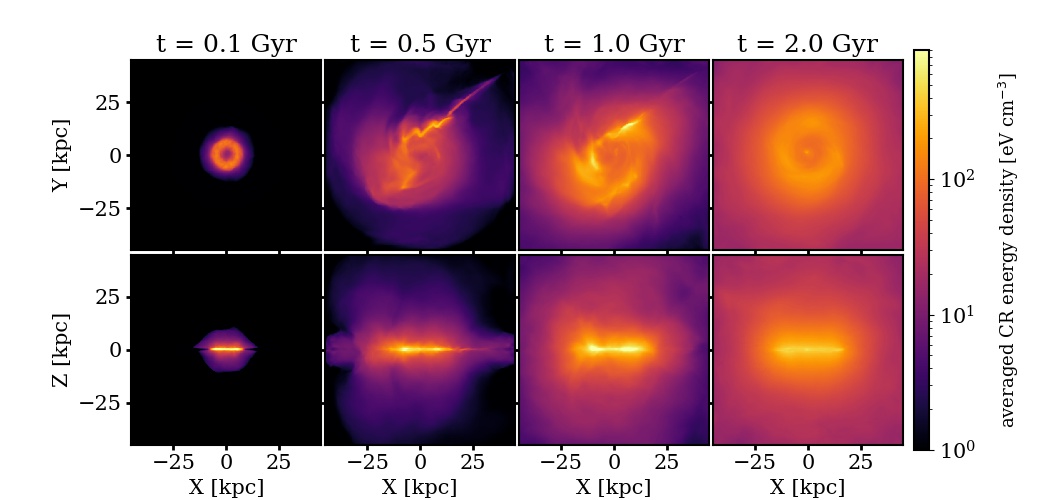}
\caption{Cosmic Ray energy density maps (averaged over the line of sight) for the p1 simulation both face-on (top panels) and edge-on (bottom panels) at t = 100~Myr, 500~Myr, 1~Gyr and 2~Gyr. As a non thermal component travelling at relativistic speeds, cosmic rays are able to quickly leave the disc and build a vertical gradient of pressure over the whole spatial domain, allowing for the disc to create gas outflows.}
  \label{evolcr}
\end{figure*}

\begin{figure}
  \includegraphics[scale=0.5]{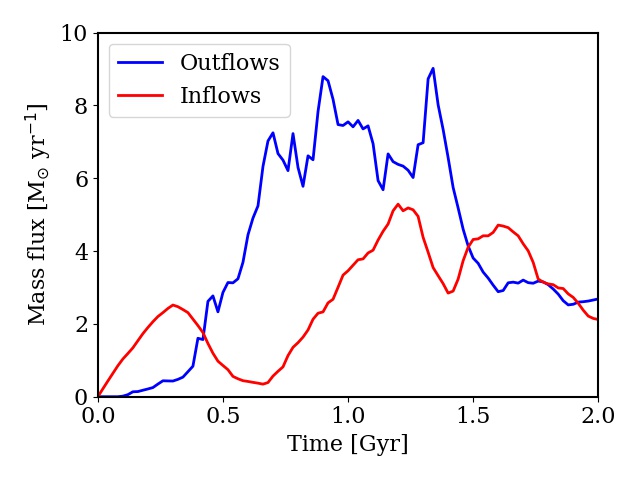}
\caption{Vertical inflow and outflow rates at 10~kpc from the disc as a function of time for the p1 simulation. The first few hundreds of Myr are dominated by accretion from the halo, however from  around 500 Myr strong outflows depart from the disc and lead to high outflow rates. The last 500~Myr show a balance between outflows and inflows.}
  \label{inf_outf}
\end{figure}

\begin{figure}
  \includegraphics[scale=0.5]{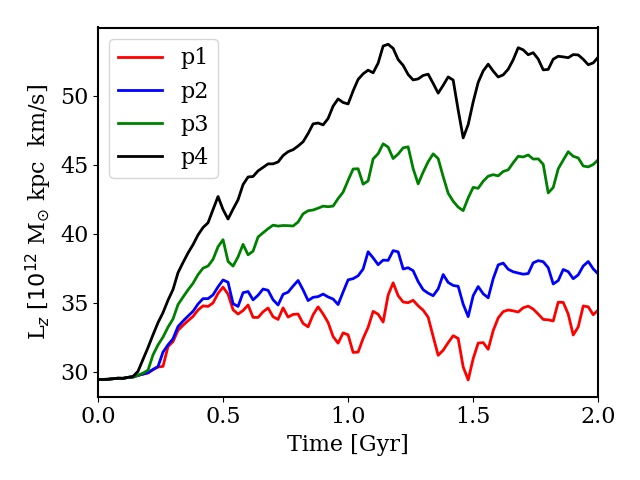}
\caption{Angular Momentum (Z-component) of the gaseous disc as a function of time, for the p1 to p4 simulations. The inflowing stream increases the disc angular momentum, and increases it more from p1 to p4 as planned. }
  \label{Lzt_comp}
\end{figure}

\begin{figure*}
  \includegraphics[trim=20 0 0 0, clip, scale=0.7]{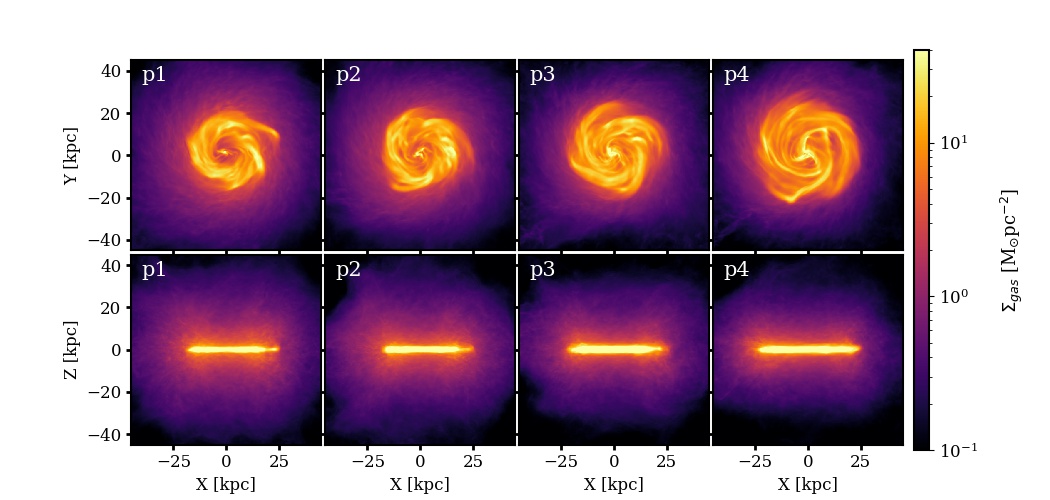}
\caption{Gas column density at t = 2~Gyr compared for p1, p2, p3 and p4 (from left to right), face-on (top panels) and edge-on (bottom panels). From p1 to p4, the disc becomes more extended and thicker due to the higher angular momentum input.}
  \label{comp_dens}
\end{figure*}

   We first study in details one simulation, p1 (lowest angular momentum input, stream in the disc plane), which will be our fiducial case to understand how CR-driven outflows are produced. Note however that the general trends and evolution presented in this section are also valid for any of our 7 other simulations.\\
   We show in Fig.~\ref{p1evol} the time evolution of projected gas density trough 6 different snapshots for p1, in face-on and edge-on view. When the stream hits the galaxy, the disc morphology is changed, creating a ring from which spiral arms will develop, which are also visible in the stellar component (Fig.~\ref{evolpart}). Note that a faint ring structure is already visible towards the beginning of the simulation before the stream hits the disc, however it is very short-lived and does not grow as strong as the one triggered by the stream (the NS simulation does not display any ring after the first few hundreds of Myr, see \href{http://piernik.umk.pl/results/2021b/}{piernik.umk.pl/results/2021b} to watch the corresponding video).
   At the end of the simulation, the galaxy shows all the morphological characteristics of a spiral galaxy, with spiral arms in the stellar and gaseous components, and with most stars concentrated in a thin ($|z| < 5$~kpc) disc. We computed the rotation curve of the stellar component and found a very realistic MW-like profile, with a maximum around 220 km s$^{-1}$ reached at $\sim$ 15~kpc, followed by a flat profile in the outer part of the disc. \\
   In the edge-on view, one sees in Fig.~\ref{p1evol} that quickly after the beginning of the simulation, as star formation sets in, gas starts leaving the disc vertically in cosmic ray driven outflows, and goes to high altitudes above and below the disc, propagating throughout the whole domain. These important outflows of gas leaving the disc for the halo were also observed in e.g. \cite{1991A&A...245...79B,2013ApJ...777L..38H,2014MNRAS.437.3312S}. We can also look at those outflows in the Z-component of the gas velocities (Fig.~\ref{evolvz}), with the area above the disc at $\sim$ 0.5~Gyr showing predominantly positive vertical velocities, and the area below the disc showing negative ones, which corresponds to gas leaving the disc. Note that at the beginning gas vertical movements are mostly due to gas accretion from the halo by the disc and are therefore directed towards the latter, and that starting $\sim$ 1~Gyr the gas that has been pushed to the halo by the outflows seems to be reaccreted by the disc, balancing the outflows. However throughout the whole simulation one can still clearly see the gas outflows from the disc, even though they seem to become more and more originated from the centre of the galaxy over time. These outflows are launched due to vertical gradients of cosmic ray pressure, as described in the introduction part, since there is no other mechanism in our simulations able to create outflows. We display this gradient of pressure in the map of cosmic ray gradient acceleration, displayed at t = 2~Gyr in Fig.~\ref{cracc}, and which we obtain by dividing the cosmic ray energy gradient (in absolute value) by the local gas density. We see that the acceleration is very strong right above and below the disc, especially close to the centre. The values obtained for the accelerations are in agreement with other works for CR generated outflows, such as \cite{2018MNRAS.479.3042G}. The CRs necessary to build this pressure gradient are emitted in star formation, which starts shortly after the simulation starts, as shown in Fig.~\ref{outfp1} (top panel). The star formation rate (SFR) increases brutally when the stream hits the disc, with a clear peak around 380~Myr after the beginning of the simulation, and remains high for about 1~Gyr (duration of the gas inflow), before dropping to very low values, shortly after the stream has been entirely absorbed by the disc. The mapping of star formation is illustrated in Fig.~\ref{p1evol}, and is present mostly in the thin disc in the spirals and at the centre. The CRs are therefore produced in the disc early on and quickly propagate in all directions to populate the whole simulation domain, which we show in Fig.~\ref{evolcr};  this allows the outflows to leave the disc alongside the CRs. \\
   We compute the flux of gas crossing different Z-planes simultaneously below and above the disc, and pointing away from the disc. We display the results in Fig.~\ref{outfp1} (top panel). Similarly to the SFR, the amount of outflows increases brutally when the stream reaches the disc and then stays high for about 1~Gyr, before dropping to lower values. The amount of outflows depends strongly on the height of the Z-plane we consider, with less outflows at higher altitudes, as one could expect. However at all altitudes the behaviour seems to be the same, and outflows follow the SFR closely, with just a small time delay increasing with altitude. This strong link between SFR and outflow rates shows the tight connection between cosmic rays emitted in the star formation feedback and outflows. At 10~kpc, we find that around 8.25 $\times 10^{9}$~$\Msun$ of gas leaves the disc to the halo in the 2~Gyr of the simulation. \\
   The mass loading factor is obtained by dividing the amount of outflows by the star formation rate, and is also displayed in Fig.~\ref{outfp1} (bottom panel). It is of the order of unity during the phase of strong star formation and outflows at 10~kpc from the disc, which is comparable to the values found in \cite{2013ApJ...777L..38H}. 
   \cite{2013ApJ...777L..16B} obtains mass loading factors of $\sim$  0.5 at 20~kpc from the disc for Milky Way-like galaxy outflows models driven by diffuse CRs, while \cite{2017ApJ...834..208R} find mass loading factors ranging from from $\sim 0.25$ to $\sim2$ at 10~kpc altitude for outflows relying on the CR streaming propagation. Those values are consistent with our results, and show the significance of the inclusion of cosmic rays in our simulations for the disc evolution. The mass loading increases towards the end of the simulation, as the result of the SFR dropping to nearly zero, while the outflows, although lower, are still important. Indeed, those outflows being launched by the vertical gradient of pressure set by cosmic rays, they remain relatively strong even after the star formation rate has nearly stopped. \\
   Since the outflowing gas seems to reach high altitudes, filling the whole computational volume, we also checked if some of the gas expelled could leave the galaxy. We computed the gas vertical velocity at the borders of the domain, and summed over time all the gas exiting the simulated volume with a vertical velocity higher than the escape velocity. This only constitutes a lower limit of the amount of gas escaping the galaxy due to these outflows, as with the gradient of pressure built by cosmic rays, the gas does not need to reach the escape velocity to leave the galaxy, and besides gas can also escape the galaxy in other directions. We find this way that over the 2~Gyr of evolution, at least $1.44 \times 10^{7}$~$\Msun$ of gas leaves the galaxy this way, showing that those outflows are able to expel gas out of the galaxy, in agreement with e.g. \cite{1991A&A...245...79B,2018ApJ...854...89M}. 
   As stated above, when looking at Fig.~\ref{evolvz}, one sees that if the halo seems dominated by outflows at early times, at later times this trend is a bit weaker, and one finds areas of gas falling back to the disc. To check this we also measured the gas inflow rate as a function of time, i.e. the gas flowing towards the disc, and plotted it in Fig.~\ref{inf_outf} for the altitude of 10~kpc, compared to the outflow rate. We can see that the inflows show a little peak at the beginning of the simulation corresponding to gas accreted from the halo, but this accretion doesn't last and after $\sim$  600~Myr the inflow rate drops to almost zero, while the outflows increase and completely dominate the vertical gas movements. However starting t $\sim$ 800~Myr the inflows rate increases again (while the outflows remain high), most likely due to the new re-accretion of the gas that has just been pushed into the halo by the outflows. This illustrates the galactic fountain mechanism: gas is ejected from the disc but is then reaccreted at later times. \\

   \subsection{Impact of the angular momentum on the outflows}
   \label{comp}

\begin{figure}
  \includegraphics[trim=20 0 0 0, clip, scale=0.5]{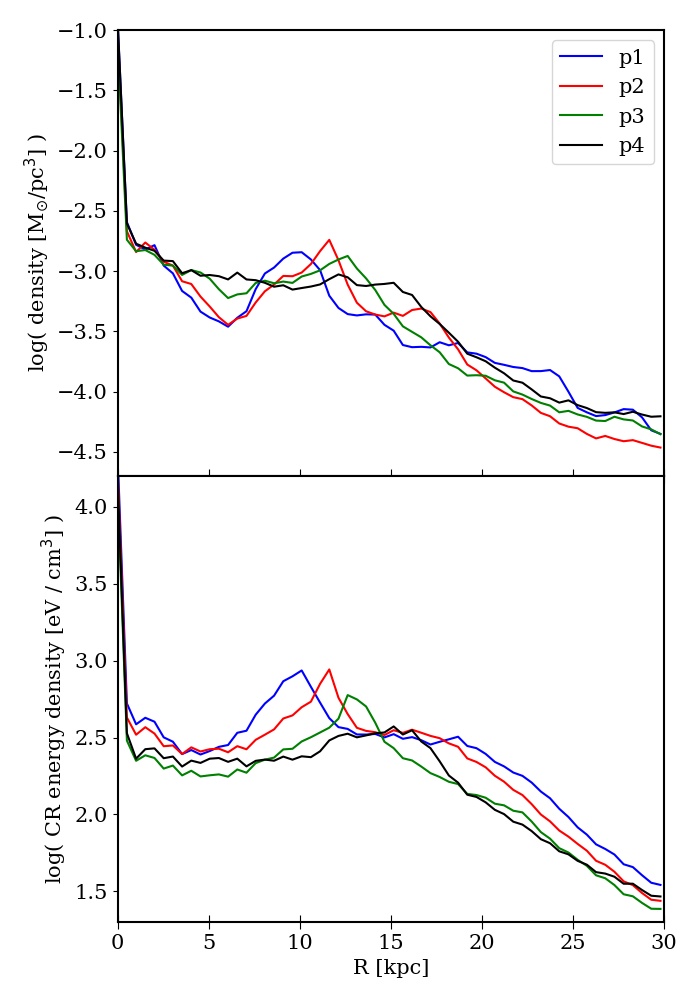}
\caption{Surface density profiles of the gas (top panel) and the CR energy (bottom panel) in the disc ($|Z|$ < 5~kpc) for the p1-p4 simulations at t = 1~Gyr.}
  \label{1dprof}
\end{figure}

\begin{figure}
  \includegraphics[scale=0.5, trim=0 10 0 70, clip]{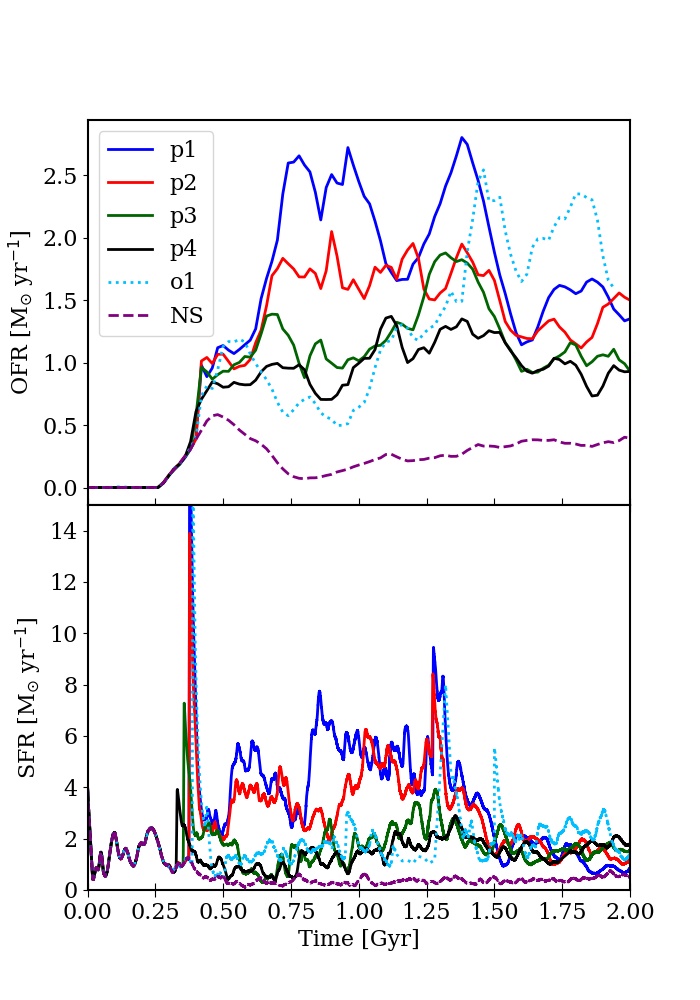}
\caption{Top panel: outflow rate as a function of time at 25~kpc altitude for p1, p2, p3, p4, o1 and NS. The low angular momentum runs have stronger outflows than the high angular momentum runs, especially during the early, strong outflowing phase around 1~Gyr. Bottom panel: total Star Formation Rate as a function of time for the same simulations. The less angular momentum injected in the disc, the stronger the star formation rate during the strong star formation phase for the p-simulations. The o1 simulation, despite having a similar angular momentum input as p1, shows significantly less star formation as well as less outflows. The NS (no stream) simulation has much lower star formation and outflows than all the other simulations.}
  \label{outf_comp}
   \end{figure}

\begin{figure}
  \includegraphics[scale=0.57, trim=15 0 0 0, clip]{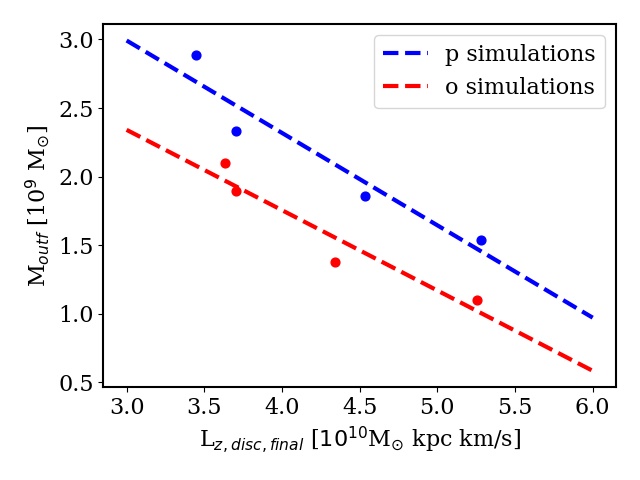}
\caption{Total outflowing mass at 25~kpc altitude over 1.6~Gyr from the moment the stream hits the disc, as a function of the final Z-component angular momentum of the disc, for our 8 simulations. Each point represents one simulation, with in blue from left to right p1, p2, p3 and p4, and in red o1, o2, o3 and o4. The dashed lines represent the linear regression of the p1-p4 simulations, and the o1-o4 simulations separately. The corresponding correlation coefficients are $r_{\rm corr}^2=0.91$ for the blue line, and 0.92 for the red one. This plot shows the strong correlation between angular momentum and the production of CR driven outflows.}
  \label{L_outf}
   \end{figure}

   In Fig.~\ref{Lzt_comp} we show the evolution of the disc angular momentum for the p-simulations, with the disc defined here as a 35~kpc cylindrical radius cylinder with 10~kpc height ($|z|<5$~kpc). As expected, the angular momentum increases more strongly for higher velocity accretion streams. \\
   Fig.~\ref{comp_dens} displays the final gas surface density distributions for each of the in-plane accretion simulations, and Fig.~\ref{1dprof} (top panel) shows the gaseous surface density profile at 1~Gyr; we see that the evolution of the morphology and of the outflows in each case shows important differences. Note that the high density peak in the very centre might be an artifact due to a large quantity of gas concentrated in only a few grid cells. As expected (see e.g. \citealt{1997ApJ...482..659D}), higher angular momentum discs are more extended. Furthermore, the low angular momentum simulations show slightly thinner discs than the high angular momentum ones, which we confirmed by looking at their vertical density profiles. In all cases, the outflows have transported gas into the halo, more efficiently though for low angular momentum accretion. To quantify this, we plot in Fig.~\ref{outf_comp} the outflow rates (top panel) as a function of time as in Fig.~\ref{outfp1}, at 25~kpc from the disc, for the p-simulations. We can see that indeed outflow rates decrease from p1 to p4, i.e. with increasing angular momentum. Accordingly, low angular momentum simulations have lower final disc masses (from 1.72$\times$~$10^{10}$~$\Msun$ for p4 to 1.29$\times 10^{10}$~$\Msun$ for p1), due to more gas leaving the disc in the outflows (the inflowing stream injects the same mass into the disc for all the simulations). Furthermore, measuring the amount of gas escaping the domain in the Z-direction as in section \ref{p1}, we find that from p1 to p4 the galaxies lose less and less gas, from 1.44 $\times 10^7~\Msun$ for p1 to 3.47 $\times 10^6$~$\Msun$ for p4 over the 2~Gyr of evolution. \\
   In order to quantify the connection of outflow and angular momentum, we plot in Fig.~\ref{L_outf} the total amount of gas that has left the disc to cross the $|Z|$ = 25~kpc planes, as a function of the final disc angular momentum. We took 25 kpc to be able to also include the o1-o4 simulations in this plot, for which any lower distance would be polluted by the inflowing stream (starting 20~kpc above the disc, and going through the disc, see Fig.~\ref{setup}). The angular momentum of the disc is computed at $t=t_{\rm hit}+1.6$~Gyr ($t_{\rm hit}$ being the time the stream hits the disc, which is different for the simulations). We find a strong correlation between the final angular momentum of the disc and the mass of gas having left the disc for the p-simulations, with a correlation coefficient $r_{\rm corr}^2=0.91$. \\

   
\begin{figure*}
  \includegraphics[scale=0.5, trim=80 0 0 0, clip]{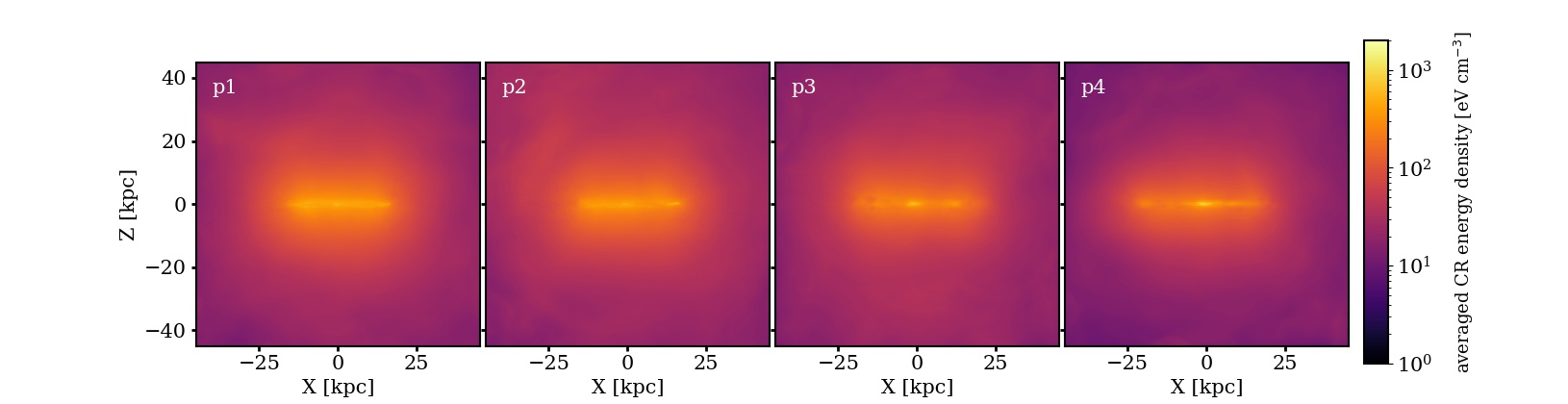}
\caption{CR energy density edge-on maps (averaged in the Y direction) for the p1 to p4 simulations (from left to right) at t = 2~Gyr. The more angular momentum the disc has, the less CRs it emitted and propagated throughout the domain.}
  \label{comp_cr1}
\end{figure*}

   To understand and interpret this correlation, we investigated what leads to more outflows in the low angular momentum fed discs. Since our outflows are CR driven, we compared the CR energy density maps at the final times (Fig.~\ref{comp_cr1}), as well as the surface density profiles of CR energy in the disc at 1~Gyr (Fig.~\ref{1dprof}, lower panel). We see that less CRs have been emitted from p1 to p4, and that they extend to lower altitudes above the disc. This will in turn decrease the outflows able to reach a large scale, and thus explains the decreasing trend observed in Fig.~\ref{L_outf}. The next step is therefore to understand why there are more cosmic rays emitted in the low AM cases than in the high AM ones. As cosmic rays in our simulations are released through supernovae explosions, we will next turn to the star formation rate. As we compare the amount of star formation in our different simulations as a function of time in Fig.~\ref{outf_comp} (bottom panel), we find that the low angular momentum simulations form more stars during the stream input, and therefore release more cosmic ray energy. This increased star formation is crucial to understand our correlation as it is the cause of the observed stronger outflows, and is likely due to a higher concentration of gas in the disc, which is more compact in the low angular momentum cases (smaller, thinner discs). Due to the power-law relation of SFR on gas density (SFR $\propto \rho^{3/2}$, which mimics the Schmidt-Kennicutt relation, CR energy injected per unit mass of thermal gas is higher for the low AM gas than for the high AM gas. This can be seen in the disc surface density profile of CR energy in Fig.\ref{1dprof} (lower panel), where more CRs are found in areas of high gaseous density.
We also added the NS (no stream) simulation to Fig.~\ref{outfp1} as a reference. The outflows (top panel) are much weaker than for the other simulations, because the star formation is also much lower (bottom panel). This is expected as the gas concentration will necessarily be lower than in the cases with gas accretion in a stream.\\

We then turn to the simulations with the stream incoming out of the plane, o1 to o4. Those simulations aim to constitutes a less idealised setup than the p1-4 simulations, as gas is unlikely to arrive exactly in the disc plane in the event of the accretion of gas from the intergalactic medium. The only difference between this set of simulation and the in-plane set is that the stream starts 20~kpc above the disc, but the initial input of angular momentum in the Z-direction through the stream is identical. The resulting simulations also produce realistic spiral galaxies, similar to the p1-p4 set, although with consistently thicker discs, due to the vertical input of matter through the stream. The stream, contrary to the p1-4 set of simulations, is not absorbed by the disc at the moment it reaches it, but goes through it, and is then reaccreted shortly after. We find overall very similar results for those simulations to the p1-p4 ones, i.e. from o1 to o4, the disc angular momentum increases, while the star formation rate decreases, leading to a decrease of the outflow rate as well. Therefore everything we said about the in-plane stream simulations remains valid for the out-of-plane stream ones, and we added those 4 simulation in the correlation plot between angular momentum and outflows in Fig.~\ref{L_outf}. A clear correlation is also present for this set of simulation ($r_{\rm corr}^2=0.92$), but is distinct from the p1-p4 set, with weaker outflows for a given disc angular momentum, which is also visible in the top panel of Fig.~\ref{outf_comp}. This is caused by a lower star formation rate in those simulations, as shown in the lower panel of Fig.~\ref{outf_comp}, which we explain by the more violent entry of the stream into and through the disc, disrupting the disc structures and increasing the velocity dispersion of the gas, making it harder to settle in dense star formation areas. Furthermore, the disc being thicker indicates a less compact disc than in the p1-4 simulations, which directly impacts the star formation. It is interesting to note that both correlation lines seem to be nearly parallel, indicating a very similar behaviour of the amount of outflows with angular momentum for both sets of simulations. For the o-simulations we compared outflows as a function of time, star formation rate, angular momentum evolution, and mass loadings as performed above for the in-plane simulations, and found very similar results to the p-simulations. Therefore, the o1-4 set of simulation confirms the trend found for the outflows with the amount of angular momentum in the disc, in a different, less idealised setup than the p1-4 set. The discs formed this way evolve differently and therefore the outflows are also different, so that they do not fit with the p1-4 correlation, however they still show very similar trends and thus remain self-consistent.\\

As noted for p1 in Section \ref{p1}, the accretion of the inflowing stream often leads to the formation of a gaseous ring structure, particularly visible for o1 and o4 in Fig.~\ref{setup}. This ring typically lasts for a few hundreds of Myr, and is found in most of our simulations (including p1-4) but is found stronger in the o simulations, as well as in the low angular momentum runs, which can also be seen in the surface density profile of the discs at 1~Gyr the p-simulations (Fig.~\ref{1dprof}, top panel), with a drop in density around 5 kpc radius higher for low AM cases. We refer the reader again to the videos of the gas evolution on the \textsc{piernik} webpage to observe the formation and evolution of those rings (\href{http://piernik.umk.pl/results/2021b/}{piernik.umk.pl/results/2021b}).

   \subsection{Angular momentum of the outflows}
   \label{windsAM}

\begin{figure}
  \includegraphics[scale=0.5, trim = 0 0 0 65, clip]{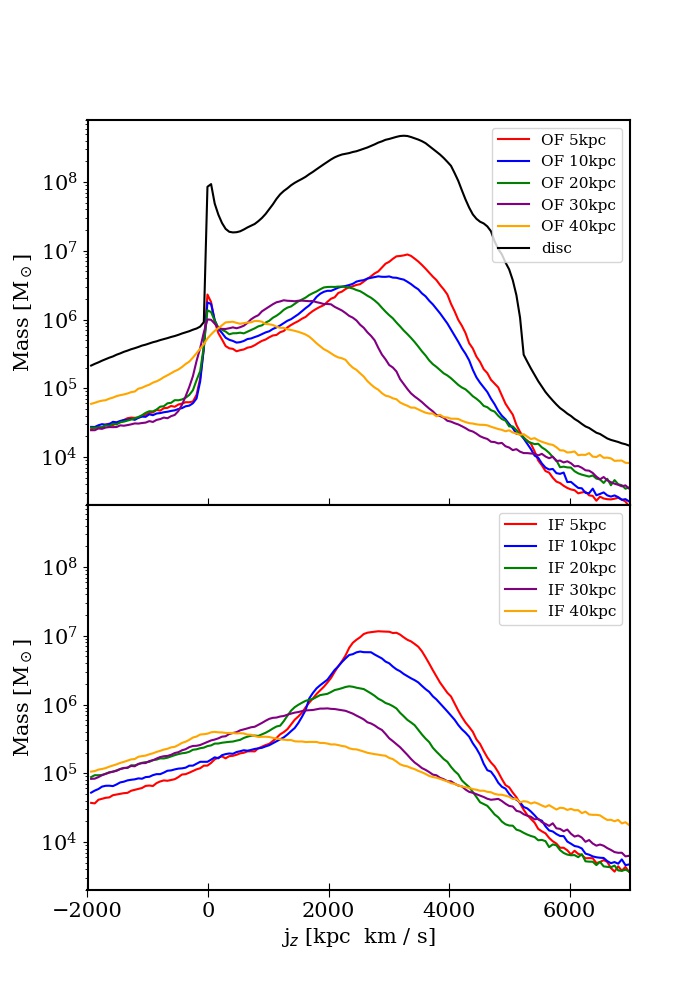}
\caption{Top panel: specific angular momentum $j_z$ distribution of the outflowing gas flux at different altitudes (colored lines) for the p1 simulation, summed in the last 200~Myr, as well as the $j_z$ distribution of the disc (in black) averaged in this same time period. Bottom panel: same for the inflowing gas. The bin size in $j_z$ is 600~kpc~km~s$^{-1}$. The specific angular momentum of the outflows decreases with altitude, suggesting stronger outflows have less angular momentum.}
\label{Lz_outf_z}
   \end{figure}

\begin{figure}
  \includegraphics[scale=0.5, trim = 0 0 0 65, clip]{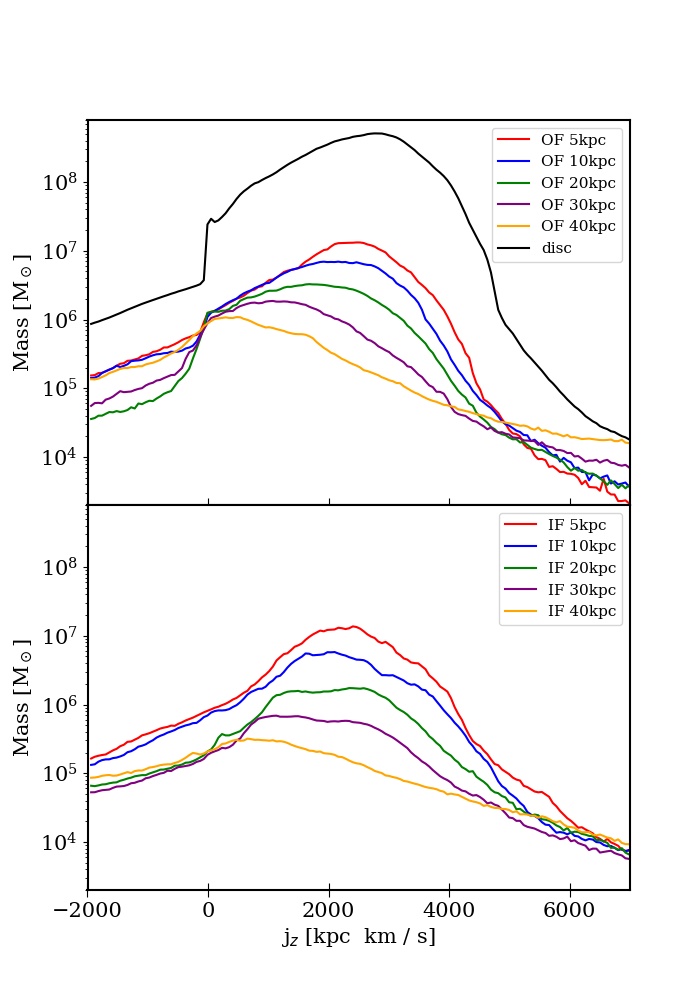}
\caption{Same figure as Fig.~\ref{Lz_outf_z}, but for the o1 simulation.}
\label{Lz_outf_z_o1}
      \end{figure}

\begin{figure}
  \includegraphics[scale=0.5]{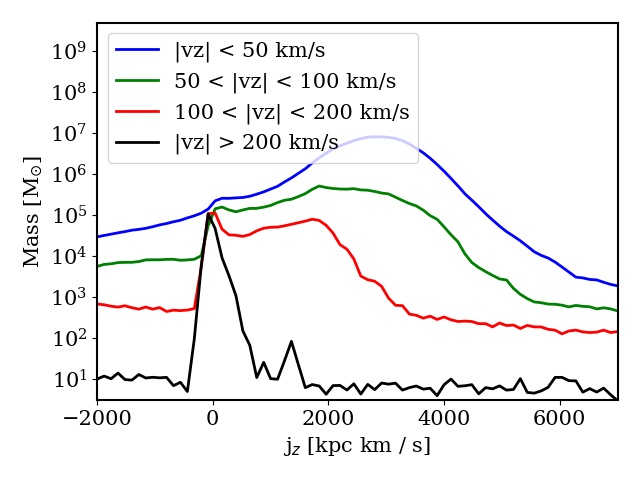}
  \includegraphics[scale=0.5]{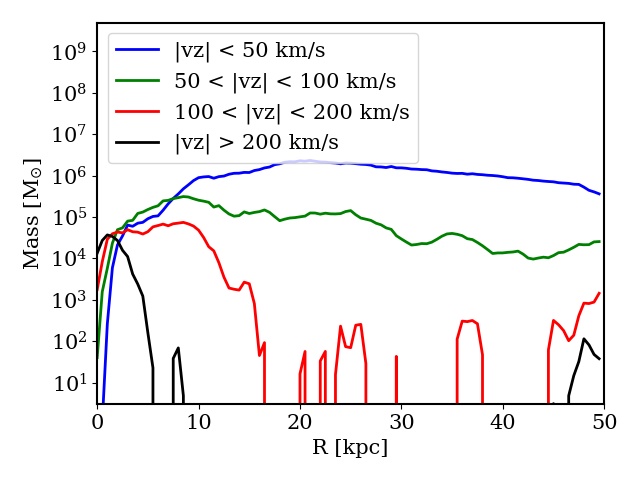}
\caption{Top panel: specific angular momentum $j_z$ distribution of the outflowing gas present at 10~kpc from the disc of p1 and averaged in the last 200~Myr of the simulation, for different vertical velocities of the gas. The bin size in $j_z$ is 600~kpc~km~$s^{-1}$. Slow outflows carry most of the high angular momentum gas, while faster outflows have much lower angular momentum. Bottom panel: radial distribution of the same outflowing gas also for different velocities, with a bin size in R of 500~pc. The fast outflows are generated from the central parts of the galaxy.}
  \label{vz_hist}
\end{figure}

 In Fig.~\ref{Lz_outf_z} and \ref{Lz_outf_z_o1} we show for p1 and o1 the specific angular momentum $j_z$ distribution of the outflowing gas at different altitudes from the disc, averaged over the last 200~Myr of the simulation. The disc specific angular momentum distribution is also shown for comparison. Both simulations show similar results: gas leaving the region close to the disc plane has a specific angular momentum distribution very similar to the disc itself, however only gas with lower angular momentum reaches high altitudes. In the bottom panel of Fig \ref{Lz_outf_z} and \ref{Lz_outf_z_o1} we show the corresponding angular momentum distribution for the inflowing gas. Close to the disc midplane the inflow angular momentum is similar to the outflow, indicating a fountain flow. There is a trend for the inflows having a slightly lower peak in the distribution and the peak of close to zero angular momentum outflow is missing in the inflows. This zero peak is more important for p1 than for o1 in the outflows, probably because of a higher gas concentration and star formation in the centre for p1 (see Fig.~\ref{setup}).\\
   To understand this trend, we look at the specific AM of outflows at 10~kpc from the disc for different vertical velocities of the gas in Fig.~\ref{vz_hist} (top panel) for p1 for the last 200~Myr of the simulation. We find that outflows carrying high AM gas are mostly slow, while fast outflows carry only low AM material, with a peak close to 0. This explains why we find mostly low AM gas at high altitudes, and why high AM gas does not leave the vicinity of the disc; the velocity of the high AM gas is just too low to reach high altitudes. Note that low angular momentum material is still found in slow outflows as well, but high angular momentum gas is not found in fast outflows. Therefore, it seems that the only gas ejected in the outflows that is able to really escape the galactic disc, its vicinity (and therefore is not reaccreted right away, as shown by the absence of the zero angular momentum peak in the inflows), and even leave the galaxy for the intergalactic medium, is low angular momentum gas. This is in agreement with the results for thermal driven outflows (\citealt{2011MNRAS.415.1051B,2014MNRAS.443.2092U,2016ApJ...824...57C}; see also \citealt{2017ARA&A..55...59N} for a review) which found that outflows were removing predominantly low angular momentum gas from galactic discs, but is shown here for the first time for cosmic ray driven outflows. This will have consequences on the disc AM distribution, as outflows permanently remove some very low AM gas from the centre, therefore likely increasing its specific AM. The fact that fast outflows eject low AM gas can be seen and understood both in Fig.~\ref{evolvz} and in the lower panel of Fig.~\ref{vz_hist} displaying the radial distribution of the outflows at different vertical velocities: the faster vertical velocities throughout the simulation are emitted from the very central part of the disc, where low AM gas is expected to be, while slow outflows are emitted at all radii except for the very centre. This is also where the cosmic ray gradient acceleration is the strongest (see Fig.~\ref{cracc}), as it is an area of strong star formation. The centre of the galaxy is therefore the region that allows to launch very fast, powerful outflows composed of low AM gas escaping the galaxy, thanks to the high SFR found there. The cosmic rays thus show to be an important way of transporting low AM gas with potentially high metallicities from the galactic centre into the more metal poor gaseous halo.\\
   Those results are also true for the other simulations, but it is interesting to note that for higher AM discs, outflows have a $j_z$ distribution more and more shifted towards higher values, so that high AM discs also seem to produce higher AM outflows. They however all show the peak around $j_z \sim0$ corresponding to outflows departing from the centre.\\

   \section{Discussion}
   \label{discussion}

We have run the simulations presented here also at a factor of two lower resolution (256$^3$) cells with similar results as presented here. Therefore at least for these resolutions, the results are convergent. At this low resolution we have also tested a stream configuration which is counter-rotating to the main disc. As expected the galactic disc loses even more angular momentum with a higher star formation rates and stronger galactic outflows. This confirms the trends with angular momentum presented above. 

We have also varied the star formation efficiency between 0.1 and 1. With increasing efficiency the outflows are, overall, more powerful. At a given efficiency, the trends with the angular momentum of the infalling stream, presented here, remain.
 We have also run a set of simulations without CR injection which have produced no outflows. 

 We ran some simulations without disabling the star formation in the inflowing gas stream, to test the robustness of our results outside the parameters of our setup. As the stream is very concentrated, star formation starts immediately in the stream, releasing cosmic ray energy which allows the gas to propagate from the stream, broadening it and forming in the end a very wide and diffuse inflow, as compared to the very narrow stream in the simulations we presented in this paper. The resulting disc galaxy has a different morphology as the violence of the stream impact is considerably lower, and it therefore shows a smoother and earlier type spiral structure, even for the o1-4 simulations. However the trend of stronger outflows for lower angular momentum input is still present in this configuration, showing again that our results remain valid independently of our choice of parameters and of our setup. We ran 8 versions of those simulations, with the same stream parameters as the ones presented in this paper, and found an interesting result: for the correlation plot of Fig.~\ref{L_outf}, the 8 simulations end up on the same correlation line, i.e. there is no clear difference between the p1-4 and o1-4 sets of simulations for this trend when star formation is enabled in the stream. This supports the claim that the differences between the p1-4 and o1-4 sets shown in this paper are indeed due to the impact of the stream, much more violent for o1-4 when star formation is disabled in it, and producing more disrupted discs with lower SFR and therefore less outflows.

We tried to make our correlation plot (Fig.~\ref{L_outf}) for different distances from the disc (including 10~kpc for the p1-4 simulations), and despite obtaining different values for the outflowing masses (as seen in Fig.~\ref{outfp1} the amount of outflows decreases with altitude), the correlation remains, always with a high value of $r_{\rm corr}^2$. 

To isolate the cosmic ray driven outflows, our simulations did not include any thermal star formation feedback, and therefore ignored thermally-driven outflows. By adding the latter to cosmic ray driven ones, we expect even stronger outflows if thermal feedback was also included, and similar results in terms of angular momentum dependency as in low angular momentum galaxies there is more star formation. However, contrary to CR driven outflows, outflows created by pure gas heating send very little gas to high altitudes above the disc, as the gas is rapidly slowed down by ram pressure. This has been shown in previous simulations, e.g. in \cite{2016ApJ...816L..19G,2018MNRAS.479.3042G}, where at low altitudes there is an approximate equilibrium between CRs and thermal pressure, while at higher altitudes CRs are the dominant gas acceleration mechanism. Therefore by including thermal heating on top of cosmic rays, we expect stronger outflows close to the disc, but no significant differences at high altitudes, so that our results at 25~kpc from the disc as in Fig.~\ref{L_outf} should not be strongly affected.

The differences observed between the p1-4 and o1-4 set of simulations, in particular for the correlation between the outflows strength and the disc AM (Fig.~\ref{L_outf}), indicate that angular momentum is not the only parameter that comes into play when looking at outflows strengths. Here we explained those differences by the impact of the stream, leading to a different type of disc, thicker and less compact, but there might be many other factors influencing outflows or the SFR, such as the amount of gas available in the galaxy for star formation, interactions with other galaxies, and in general the way the disc has formed and evolved. Therefore we do not expect to be able to predict outflows strengths just from the AM of a galactic disc, however in this paper we have shown that there is a robust trend of higher angular momentum galaxies producing less CR driven outflows due to a lower SFR, all other things remaining equal.

Looking at the evolution of magnetic field in our simulations, we observe a gradual amplification of the magnetic field over time, as expected. The toroidal component follows the disc structures closely, in particular the spiral arms, and remains weak outside the disc. Comparing the magnetic field between our set of simulations, we find that the more angular momentum, the stronger the magnetic field in the disc for the toroidal component. We explain this by a higher turbulent energy brought to the disc by higher angular momentum streams, which have more kinetic energy. We also observe that magnetic field leaves the galaxy in the outflows, by measuring the Z-component of the Poynting flux for cells of gas escaping the computational domain with velocities higher than the escape velocity. Integrated over time and over the vertical domain boundary surfaces, we find that 1.79 $\times$ 10$^{9}$~G$^2$~pc$^{3}$ escapes the galaxy for p1 during the whole simulation, and this values decreases for higher angular momentum runs, higher angular momentum discs losing less magnetic field through outflows. As mentioned in section \ref{p1}, this value of escaping magnetic field energy is only a lower limit, as more gas and more magnetic field is expected to leave the galaxy due to the CR gradient of pressure.

\section{Summary and Conclusions} 
\label{ccl}
 
In this paper, we used the \textsc{piernik} code to create MHD + Nbody + Cosmic Rays simulations of disc galaxies accreting gas streams with different angular momenta, to study the creation of CR driven outflows.\\
The stellar disc and the dark matter halo are built from \textsc{buildgal}, and the gaseous disc is then built in equilibrium with those components, with an initial toroidal magnetic field of $\sim$ 10~$\mu$G. Cosmic Rays are then injected as the form of energy density throughout the simulation, as the only source of star formation feedback. The stream takes the form of an external inflow of gas for 1~Gyr from the side of our domain box; we run 8 simulations for 2~Gyr varying the parameters of this stream, in order to inject different amounts of angular momentum into the disc. This gas accretion represents the stream growth mechanism of galaxies in a cosmological context (\citealt{2009Natur.457..451D}), and triggers an enhanced star formation episode, as well as the formation of galactic rings (e.g. \citealt{2015MNRAS.449.2087D}). Those rings are found in most of our simulations, but are stronger for the accretion of a stream out of the disc plane, and with lower angular momentum. \\
The implementation of cosmic rays in star formation feedback, travelling through anisotropic diffusion, creates a gradient of pressure which allows the gas to leave the galactic disc (e.g. \citealt{1993A&A...269...54B, 2008ApJ...674..258E,2014ApJ...797L..18S}), launching strong outflows reaching high altitudes below and above the disc. Gas is even able to escape the galaxy this way. We compared those outflows in our 8 simulations, and found that there is a direct correlation between the amount of angular momentum accreted by the disc from the stream, and the amount of outflows: low angular momentum accretion leads to stronger outflows reaching higher altitudes than high angular momentum accretion. The star formation rate plays a central role to explain this correlation, as it is higher in low angular momentum cases, creating more cosmic rays necessary for those outflows to be launched. The higher star formation is likely to be explained by higher gas concentration in the disc, which is smaller and more compact in the case of a low angular momentum stream. The correlation, although still present, is different for simulations where the gas stream arrives in the disc plane and for the ones where it arrives above the disc. We explain this difference by the difference in the evolution of the disc; the violence of the impact of an out-of-plane inflow of gas will lead to a thicker, less compact and more disrupted disc, which implies less star formation and thus less outflows, for a similar input of vertical angular momentum. \\
Looking at the gas composing those outflows, we discover a characteristic structure in angular momentum and velocity. The gas escaping the galaxy with high vertical speeds (> 200 km/s) has mostly very low angular momentum, and is able to reach high altitudes (40~kpc and above) and therefore penetrate deep into the halo. These powerful outflows are mostly launched from the centre of the galaxy, due to the high star formation rate found in this region. On the opposite, gas outflowing at lower velocities (< 50 km/s) displays an angular momentum distribution closer to the disc one, and stays in the vicinity of the disc to be re-accreted at later times in a galactic fountain mechanism. Those slower outflows contain the high angular momentum gas, which is therefore not able to penetrate the halo. Cosmic ray driven outflows therefore tend to remove predominantly low angular momentum gas from the disc, similarly to thermal generated outflows (\citealt{2011MNRAS.415.1051B,2014MNRAS.443.2092U}). \\
Both in low and high angular momentum cases though, we find that with the parameters chosen here, cosmic ray driven outflows are strong with mass loadings values above one, and have an important impact on the galaxy evolution, launching large amounts of gas (with likely higher metallicities) into the halo, creating galactic fountains, and removing gas from the centre predominantly. We therefore present strong evidences that cosmic rays, together with magnetic field, are important components to be taken into account in galaxy simulations, in agreement with recent studies focusing on cosmic ray driven outflows (e.g. \citealt{2012MNRAS.423.2374U,2013ApJ...777L..16B,2016ApJ...816L..19G,2020MNRAS.492.3465H}).

\vspace{1cm}
\noindent \textbf{AKNOWLEDGEMENTS} \\

\noindent We thank the anonymous referee for useful comments and suggestions to improve the quality and scope of the paper. This work was supported by the Polish National Science Center under grant 2018/28/C/ST9/00443.
Thorsten Naab acknowledges support by the Excellence Cluster ORIGINS which is funded by the DFG (German research foundation) under Germany’s Excellence Strategy – EXC-2094 – 390783311.

\vspace{1cm}
\noindent \textbf{DATA AVAILABILITY} \\

\noindent  The data underlying this article will be shared on reasonable request to the corresponding author.

\bibliographystyle{mnras}
\bibliography{biblio}

\end{document}